# Algorithmic Aspects of Several Data Transfer Service Optimization Problems

**Mugurel Ionut Andreica, Politehnica University of Bucharest**
**Ion Pargaru, Politehnica University of Bucharest**
**Florin Ionescu, Konstanz University of Applied Sciences**
**Cristina Teodora Andreica, Commercial Academy Satu Mare**

**Abstract:** *Optimized data transfer services are highly demanded nowadays, due to the large amounts of data which are frequently being produced and accessed. In this paper we consider several data transfer service optimization problems (optimal server location in path networks, optimal packet sequencing and minimum makespan packet scheduling), for which we provide novel algorithmic solutions.*

## 1. Introduction

   Durable economic development under unstable economic conditions can only occur if a clear cost advantage is offered to the target customers. Among other things, this implies cutting production and maintenance costs and improving the profit margin by carefully scheduling the economic activities. In this paper we consider the particular case of (electronic) data transfer service providers. Although at this moment there is no difference between Internet service providers (ISPs) and data transfer service providers, we argue that the business need for providers of optimized data transfer services is on the rise. Since they are still an emergent economic category, in order to sustain durable economic development, the provided data transfer services will need to be highly optimized. The focus of this paper is on algorithmic aspects of data transfer service optimization. We consider server location problems (Section 2), a packet sequencing problem (Section 3), and a packet scheduling problem (Section 4). Finally, in Section 5 we conclude and discuss related work.

## 2. Connected K-Centers and K-Medians in Path Networks

Let's assume that we have $N$ network nodes arranged on a path. Each node $i$ ($1 \leq i \leq N$) has an associated coordinate $x(i)$ (its position from the beginning of the path) and a weight $w(i)$; $x(i) \leq x(i+1)$ ($1 \leq i \leq N-1$). We want to place $K \leq N$ data storage servers on $N$ consecutive nodes, such that the maximum weighted distance from a node (sum of the weighted distances from the nodes) to its (their) closest storage server(s) is minimum. The weighted distance from a node $u$ to a node $v$ is $w(u) \cdot |x(v)-x(u)|$. The two objectives define the connected K-center and the connected K-median problems in path networks. We will present linear ($O(N)$) time algorithms for these problems.

The connected K-center problem is similar to the Interval 1-Center problem presented in [1]. The Interval 1-Center problem asks for the optimal location of an interval of fixed length $L$, such that the maximum weighted distance from any of the nodes to the interval is minimum (the nodes inside the interval have distance *0*). Our problem consists of finding an interval *[q, q+K-1]* of $K$ consecutive nodes $i$ ($q \leq i \leq q+K-1$), such that the maximum weighted distance $w(j) \cdot min\{|x(a)-x(j)|, |x(b)-x(j)|\}$ for the nodes $j$ outside the interval is minimum. We start like in [1], by computing the upper envelopes of the left-oriented and right-oriented half-lines associated to every node. These consist of two sets of intervals, *[a(i),a(i+1)]* and *[b(j),b(j+1)]*. Each segment *(xa,ya)-(xb,yb)* of the upper envelope is a portion of a half-line $i$ such that its values $y_i(x)$ are larger than the values of all the other half-lines on the interval *[xa,xb]* (in [1], the last part was improperly formulated). Then, we will consider every possible interval *[q, q+K-1]*, in increasing order of $q$ ($1 \leq q \leq N-K+1$). For every such interval, we will maintain the interval *[a(u),a(u+1)]* of the upper-envelope of the right oriented half-lines, where the coordinate $x(q)$ is located, and the interval *[b(v),b(v+1)]* of the upper-envelope of the left oriented half-lines, where the coordinate $x(q+K-1)$ is located. By knowing these intervals, we can compute in

$O(1)$ time the largest weighted distance $wd_1$ from a node $j$ with a coordinate smaller than $x(q)$ to the node $q$, and the largest weighted distance $wd_2$ from a node with a coordinate larger than $x(q+K-1)$ to node $q+K-1$ (i.e. the y-values of the corresponding upper envelopes at the coordinates $x(q)$, and $x(q+K-1)$, respectively). The maximum weighted distance in the case of the interval of nodes *[q,q+K-1]* is *max{wd$_1$, wd$_2$}*. Obviously, we will choose the interval with the smallest maximum weighted distance. When sliding from the interval *[q, q+K-1]* to *[q+1, q+K]*, we will adjust the indices *u* and *v*: *(1)* while $x(q+1)>a(u+1)$ we set *u=u+1* ; *(2)* while $x(q+K)>b(v+1)$ we set *v=v+1*. For the first interval *[1,K]* we can search the indices *u* and *v* linearly. The overall time complexity is *O(N)*. For the connected K-median case, the solution is simpler. We will consider again every interval *[q, q+K-1]* ($1 \leq q \leq N-K+1$) in increasing order of *q*. We will maintain *4* values: *wleft, wsumleft, wright* and *wsumright*. Initially (for *q=1*), *wleft=wsumleft=0*, *wright* is the sum of the weights of the nodes *j* ($K+1 \leq j \leq N$) and *wsumright* is the sum of the weighted distances from the nodes *j* ($K+1 \leq j \leq N$) to node *K*. For every value of *q*, we will compute *wsum(q)=wsumleft+wsumright*. When we slide from an interval *[q, q+K-1]* to the next interval *[q+1, q+K]*, we will adjust the *4* values. We have: *(1) wleft=wleft+w(q)* ; *(2) wsumleft=wsumleft+wleft·(x(q+1)-x(q))* ; *(3) wsumright=wsumright-wright·(x(q+K)-x(q+K-1))* ; *(4) wright=wright-w(q+K)*. The optimal interval *[q,q+K-1]* is the one with the smallest value *wsum(q)*. The algorithm is linear in this case, too.

Since [1] is an important reference for the problems discussed in this section, we feel that it is appropriate to clarify some small errors which occurred in [1]. At the beginning of Subsection III.B (where the discrete 1-centers and the diameter of a cactus graph are computed), [1] introduces the notations $l_1(i)$, $l_2(i)$ and $l_3(i)$, which were incorrectly denoted by $l_1(x)$, $l_2(x)$ and $l_3(x)$. In Section VII, [1] presents a generic solution for some restricted cases of the connected K-center and K-median problems in trees. They use an algorithmic framework also introduced in [1], where the functions *UpdateAdd* and *UpdateRemove* have to be defined. In the *UpdateRemove* function, if the edge *(r,r')* is in *HS*, then we have to remove it from

HS and insert it into HT (considering its new, updated cost) ; otherwise, if HS is not empty, then we remove the edge with the smallest weight from HS and insert it into HT. In the *UpdateAdd* function we need to remove from HT the edge with the largest weight and add it to HS (only for the case $K\geq 2$). The descriptions of the *UpdateRemove* and *UpdateAdd* functions given in [1] contain some slight mistakes. In the same section, *esum* is defined as the sum of the weights of all the edges in HT. *esum* is increased (decreased) whenever an edge is added to (removed from) HT by the weight of that edge at the moment when it is (was) inserted in HT (this weight may be different from the current weight of the edge in some cases).

## 3. Optimal Packet Sequencing

We have a sequence of $N$ packets which need to be sent, in order, from a source to a destination. Every packet $i$ (the $i^{th}$ packet in the original order) has an integer type $type(i)$ ($1\leq i\leq N$; $1\leq type(i)\leq T$). When the destination receives a packet of type $q$ after a packet of type $p$, it takes $d(p,q)$ time units to decode the packet of type $q$ ($d(p,q)$ is known for every ordered pair of packets $(p,q)$; $1\leq p\leq T$; $1\leq q\leq T$). We have $K$ ($0\leq K\leq N/2$) special pairs of packets $(a,b)$ ($1\leq a\leq N$; $1\leq b\leq N$; $a\neq b$) whose order in the sequence can be swapped. Any packet $i$ ($1\leq i\leq N$) belongs to at most one special pair and any two special pairs $(a,b)$ and $(c,d)$, such that $a<b$ and $c<d$, have one of the following properties: *(1) (a<c)* and *(d<b)* ; *(2) (c<a)* and *(b<d)* ; *(3) (b<c)* ; *(4) (d<a)*. These *4* properties imply that either one of the two intervals *[a,b]* or *[c,d]* is fully included in the other, or they are totally disjoint. We want to swap the order of some special pairs of packets, such that the total decoding time of the receiver is minimum (maximum). The total decoding time is equal to the sum of decoding times of every packet $i$ ($2\leq i\leq N$), which depends on the type of the current packet $i$ and that of the previously received packet $i-1$. The first packet can be decoded in zero time (e.g. it may be a control packet and/or it may belong to no special pair). We will present a dynamic programming solution for this problem. We will maintain a stack *Stk*, where the topmost level will be given by the variable *top*.

Every position *Stk[l]* of the stack will store several fields: *Stk[l].a* and *Stk[l].b* will represent an interval of packets for which the optimum decoding time will be computed; *Stk[l].state* will be the state of the current entry in the stack – the states can be of *3* types: *advance*, *waiting$_1$* and *waiting$_2$*; *Stk[l].rez[p,q]* will be the optimum decoding time if the packet on position *Stk[l].a* was swapped (*p=1*) or not (*p=0*) and if the packet on position *Stk[l].b* was swapped (*q=1*) or not (*q=0*). For every special pair of packets *(a,b)* we define *C[a]=b* and *C[b]=a*; for every packet *i* which does not belong to any special pair, we set *C[i]=i*. We also use *d(0,\*)=C[0]= type(0)=0*.

We will start with *top=1* and *Stk[top].(a=1, b=N, state=advance, rez[0,0]=0, rez[0,1]=rez[1,0]=rez[1,1]=BadValue)*. We define *opt=min* for the minimum decoding time and *opt=max* for the maximum decoding time. If *opt=min* then *BadValue=+∞*, otherwise *BadValue=-∞*. While (*top≥1*) we will perform the following actions at every iteration of the *while* loop: *(1)* if *(Stk[top].a>Stk[top].b)* then *top=top-1* ; *(2)* otherwise, if *(Stk[top].state=advance)* then: *(2.1)* if *(Stk[top].a=Stk[top].b)* then: *(2.1.1) Stk[top].(rez[0,0]= rez[1,1]=0, rez[0,1]=rez[1,0]=BadValue)* ; *(2.1.2) top=top-1* ; *(2.2)* otherwise, if *(C[Stk[top].a]=Stk[top].b)* then: *(2.2.1)* if *(Stk[top].a+1=Stk[top].b)* then: *(2.2.1.1) Stk[top].(rez[0,0]= d(type(Stk[top].a), type(Stk[top].b)), rez[1,1]=d(type(Stk[top].b), type(Stk[top].a)), rez[0,1]=rez[1,0]=BadValue)* ; *(2.2.1.2) top=top-1* ; *(2.2.2)* otherwise: *(2.2.2.1) Stk[top].state=waiting$_1$* ; *(2.2.2.2) top=top+1* ; *(2.2.2.3) Stk[top].(a=Stk[top-1].a+1, b=Stk[top-1].b-1, state=advance, rez[0,0]=0, rez[0,1]=rez[1,0]=rez[1,1]=BadValue)* ; *(2.3)* otherwise: *(2.3.1) Stk[top].state=waiting$_2$* ; *(2.3.2) top=top+1* ; *(2.3.3) Stk[top].(a=Stk[top-1].a, b=C[Stk[top-1].a], state= advance, rez[0,0]=0, rez[0,1]=rez[1,0]=rez[1,1]=BadValue)* ; *(3)* otherwise, if (*Stk[top].state=waiting$_1$*) then: *(3.1) Stk[top].(rez[0,0]= rez[0,1]=rez[1,0]=rez[1,1]=BadValue)* ; *(3.2)* for *p=0* to *1* do: for *q=0* to *1* do: *(3.2.1)* if (*p=0*) then *Caux=d(type(Stk[top].a), type(Stk[top+1].a))* else *Caux=d(type(Stk[top].a), type(C[Stk[top+ 1].a]))* ; *(3.2.2) Caux=Caux+Stk[top+1].rez[p,q]* ; *(3.2.3)* if (*q=0*) then *Caux=Caux+d(type(Stk[top+1].b), type(Stk[top].b))* else *Caux=Caux+d(type(C[Stk[top+1].b]), type(Stk[top].b))* ; *(3.2.4)*

*Stk[top].rez[0,0]=opt(Caux, Stk[top].rez[0,0])* ; *(3.3)* for *p=0* to *1* do: for *q=0* to *1* do: *(3.3.1)* if *(p=0)* then *Caux=d(type(Stk[top].b), type(Stk[top+1].a))* else *Caux=d(type(Stk[top].b), type(C[Stk[top+1].a]))* ; *(3.3.2) Caux=Caux+Stk[top+1].rez[p,q]* ; *(3.3.3)* if *(q=0)* then *Caux=Caux+d(type(Stk[top+1].b), type(Stk[top].a))* else *Caux=Caux+d(type(C[Stk[top+1].b]), type(Stk[top].a))* ; *(3.3.4) Stk[top].rez[1,1]=opt(Caux, Stk[top].rez[1,1])* ; *(4)* otherwise (we have that *Stk[top].state=waiting$_2$*): *(4.1) rezaux[0,0]= rezaux[0,1]=rezaux[1,0]=rezaux[1,1]=BadValue* ; *(4.2)* for *p=0* to *1* do: for *q=0* to *1* do: for *r=0* to *1* do: for *s=0* to *1* do: *(4.2.1)* if *(Stk[top+1].a=Stk[top].a)* and *(p≠r* or *q≠s)* then skip over this iteration of the *for* loops ; *(4.2.2)* if *(q=0)* then *tq=type(Stk[top+1].a-1)* else *tq=type(C[Stk[top+1].a-1])* ; *(4.2.3)* if *(r=0)* then *tr=type(Stk[top+1].a)* else *tr=type(C[Stk[top+1].a])* ; *(4.2.4)* if *(Stk[top+1].a>Stk[top].a)* then *Caux=Stk[top].rez[p,q]+d(tq,tr)+ Stk[top+1].rez[r,s]* else *Caux=Stk[top+1].rez[r,s]* ; *(4.2.5) rezaux[p,s]=opt(Caux, rezaux[p,s])* ; *(4.3)* for *p=0* to *1* do: for *q=0* to *1* do: *Stk[top].rez[p,q]=rezaux[p,q]* ; *(4.4)* if *(Stk[top+1].b= Stk[top].b)* then *top=top-1* else: *(4.4.1) Stk[top].state=waiting$_2$* ; *(4.4.2) top=top+1* ; *(4.4.3) Stk[top].a=Stk[top].b+1* ; *(4.4.4) Stk[top].(b=Stk[top-1].b, state=advance, rez[0,0]=0, rez[0,1]= rez[1,0]=rez[1,1]=BadValue)*. At the end of the algorithm, the optimum (minimum or maximum) decoding time is *opt{ Stk[1].rez[p,q] | 0≤p≤1, 0≤q≤1}*. The time complexity of the algorithm is $O(N)$ (for $O(T)<O(N^{1/2})$; e.g. when the number of packet types $T$ is constant). The memory complexity is $O(N+T^2)$. Since the memory complexity cannot be larger than the time complexity (we need at least enough time to populate the memory locations), the time complexity becomes $O(T^2)$ for $O(T)>O(N^{1/2})$. The algorithm can also be expressed in a recursive manner. Let's assume that *Solve(a,b)* returns the optimum costs for the interval of packets *[a,b]*, for the cases when the packets on positions *a* and *b* are swapped or not. *Solve(a,b≤a)* is trivial. For *b>a* we have several cases. If *C[a]=a*, then *Solve(a,b)* is obtained by calling *Solve(a+1, b)* and then adding the packet on the position *a* at the front. If *C[a]=b*, then we first call *Solve(a+1, b-1)* and then we combine the results by adding a packet from the set *{a,b}* at the front and the other one at the end (swapped or not). If *C[a]<b* then we first call *Solve(a, C[a])*, then we call *Solve(C[a]+1, b)* and then we combine the results of the 2 calls.

## 4. Minimum Makespan Packet Scheduling over Multiple Disjoint Paths with Connection Initiation Times

We have a communication flow which consists of *N* identical packets (which do not necessarily have to be sent in order). The packets must be sent from the source to the destination, using some of the *P* disjoint paths available. Each path *i* ($1 \leq i \leq P$) has two parameters: a connection initiation time $CI(i) \geq 0$ and a packet sending time $PS(i) \geq 0$. If we want to send *k>0* packets on path *i*, this will take $CI(i)+k \cdot PS(i)$ time units (for *k=0*, it takes *0* time units). We want to distribute the packets over the *P* paths, such that the maximum time moment at which a packet arrives at the destination is minimum (i.e. we want to minimize the makespan of the packet distribution strategy). We will first present a solution with $O(N \cdot log(P))$ time complexity. We will maintain a min-heap *H* in which we insert every path *i*, with an initial key *Key(i)=CI(i)*. We will repeatedly extract from *H* the minimum key *N* times. Let's assume that we extracted the value *Key(i)*, assigned to path *i*. We will send a packet on path *i* at time moment *Key(i)*, which will reach the destination at time *Key(i)+ PS(i)*. Then, we remove *Key(i)* from *H*, set *Key(i)=Key(i)+PS(i)* and then re-insert *Key(i)* (the new value) into *H* (after every insertion, we also maintain the path *i* associated to every value *Key(i)*).

We now consider the restriction that only at most *Q* of the *P* available paths can be used for sending the packets. This case is identical to the previous one when *Q=P*. We will present a solution with $O(sort(P) \cdot log(TMAX))$ time complexity, where *TMAX* is the maximum possible value of the makespan and *sort(P)* is the time complexity to sort *P* numbers. We will binary search the makespan. Let's assume that we selected a value *T* within the binary search. For each path *i* ($1 \leq i \leq P$) we will compute a value *np(i)*=the number of packets which can be sent on path *i* using at most *T* time units: if *CI(i)>T* then *np(i)=0*; otherwise, *np(i)=(T-CI(i)) div PS(i)* (*A div B* denotes the integer division of *A* at *B*). We then sort the paths such that $np(path(1)) \geq np(path(2)) \geq \ldots \geq np(path(P))$, where *path(1), …, path(P)* is a permutation of the *P* paths. Out of these, we will select the first *Q* paths and compute $sumnp = np(path(1)) + \ldots + np(path(Q))$. If $sumnp \geq N$ then *T* is a feasible value for the makespan and, thus, we will test a smaller value next in the binary search; if, however, $sumnp < N$, then *T* is too small and we need to test a larger value next in the binary search. *sort(P)* may be $O(P \cdot log(P))$, or, if the values *np(i)* are small integer numbers, then *sort(P)* may be $O(P+VMAX)$, where *VMAX* is the largest possible value of *np(i)* (we can sort these values by using a procedure similar to count sort).

## 5. Conclusions

In this paper we presented several algorithmic solutions for some data transfer service optimization problems. (K-)center and (K-)median problems related to those discussed in Section 2 were studied in [1]. Data transfer scheduling problems for several classes of graphs were discussed in [4]. Minimum makespan data transfer scheduling problems with various constraints were studied in [2, 3].